\documentclass[journal,transmag]{IEEEtran}
\usepackage{booktabs}
\usepackage{soul,color}
\usepackage{multirow}
\usepackage{algorithm,algorithmic}
\usepackage{bm}
\usepackage{ifpdf}
\usepackage{cite}

\usepackage{amssymb}
\usepackage{pifont}
\ifCLASSINFOpdf
  \usepackage[pdftex]{graphicx}
\else
  \usepackage[dvips]{graphicx}
\fi

\usepackage{amsmath}
\usepackage{array}

\ifCLASSOPTIONcompsoc
  \usepackage[caption=false,font=normalsize,labelfont=sf,textfont=sf]{subfig}
\else
  \usepackage[caption=false,font=footnotesize]{subfig}
\fi
\usepackage{url}

\begin{document}

\title{Current Induced Dynamics of Multiple Skyrmions with Domain Wall Pair and Skyrmion-based Majority Gate Design }

\author{\IEEEauthorblockN{Zhezhi He\IEEEauthorrefmark{1},~\IEEEmembership{Student Member,~IEEE},
Shaahin Angizi\IEEEauthorrefmark{1}, ~\IEEEmembership{Student Member,~IEEE},
Deliang Fan\IEEEauthorrefmark{1},~\IEEEmembership{Member,~IEEE}}
\IEEEauthorblockA{\IEEEauthorrefmark{1}Department of Electrical and Computer Engineering,
University of Central Florida, Orlando, 32816 USA}

\thanks{Corresponding author: Deliang Fan (email: dfan@ucf.edu) }}

\markboth{}%
{Shell \MakeLowercase{\textit{et al.}}: Bare Demo of IEEEtran.cls for IEEE Transactions on Magnetics Journals}

\IEEEtitleabstractindextext{%
\begin{abstract}
As an intriguing ultra-small particle-like magnetic texture, skyrmion has attracted lots of research interests in next-generation ultra-dense and low power magnetic memory/logic designs. Previous studies have demonstrated a single skyrmion-domain wall pair collision in a specially designed magnetic racetrack junction. In this work, we investigate the dynamics of multiple skyrmions with domain wall pair in a magnetic racetrack. The numerical micromagnetic simulation results indicate that the domain wall pair could be pinned or depinned by the rectangular notch pinning site depending on both the number of skyrmions in the racetrack and the magnitude of driving current density. Such emergent dynamical property could be used to implement a threshold-tunable step function, in which the inputs are skyrmions and threshold could be tuned by the driving current density. The threshold-tunable step function is widely used in logic and neural network applications. We also present a three-input skyrmion-based majority logic gate design to demonstrate the potential application of such dynamic interaction of multiple skyrmions and domain wall pair.

\end{abstract}

}

\maketitle

\IEEEdisplaynontitleabstractindextext
\IEEEpeerreviewmaketitle

\section{Introduction}

Skyrmion is a particle-like spin texture which has been experimentally observed both in B20 type bulk material  (MnSi \cite{muhlbauer2009skyrmion}, Fe\textsubscript{1-x}Co\textsubscript{x}Si \cite{yu2010real,munzer2010skyrmion}, FeGe \cite{yu2011near, huang2012extended}) and ultra-thin ferromagnetic film (Ta/CoFeB/TaO\textsubscript{x} \cite{jiang2016direct} and Pt/CoFeB/MgO \cite{litzius2016skyrmion}) favored by Dzyaloshinskii-Moriya Interaction (DMI). Owing to its topological stability, skyrmion has been widely investigated in recent years for racetrack memory \cite{tomasello2014strategy}, logic \cite{zhang2015magnetic_logic} \cite{xing2016skyrmion}, oscillator \cite{zhang2015current}, transistor \cite{zhang2015magnetic_transistor} and neuromorphic computing \cite{Huang2017}. The dynamics between single domain wall pair (DWP) and single skyrmion has been studied in \cite{xing2016skyrmion}. In this work, we investigate the dynamical interaction between multiple skyrmions and single domain wall pair in a magnetic racetrack. The conducted micromagnetic simulation reveals that the pinning and depinning of a DWP from a rectangular notch is determined by both the number of skyrmions (defined as \textit{threshold skyrmion number}) and the magnitude of driving current density. Increasing the driving current density leads to larger skyrmion and DWP drifting velocity, but lowering the threshold skyrmion number. Such skyrmions and DWP interaction behavior can be used to design a threshold-tunable step function, in which the inputs are skyrmions and threshold could be tuned by the driving current density. In the end, we present a skyrmion-based majority logic gate design as an example to demonstrate a potential logic application. 

\section{Skyrmions-Domain Wall Pair Dynamics}

\subsection{Device structure and basic operations}

\begin{figure}[ht]
    \vspace{0in}
	\centering
	\includegraphics[width=0.47\textwidth]{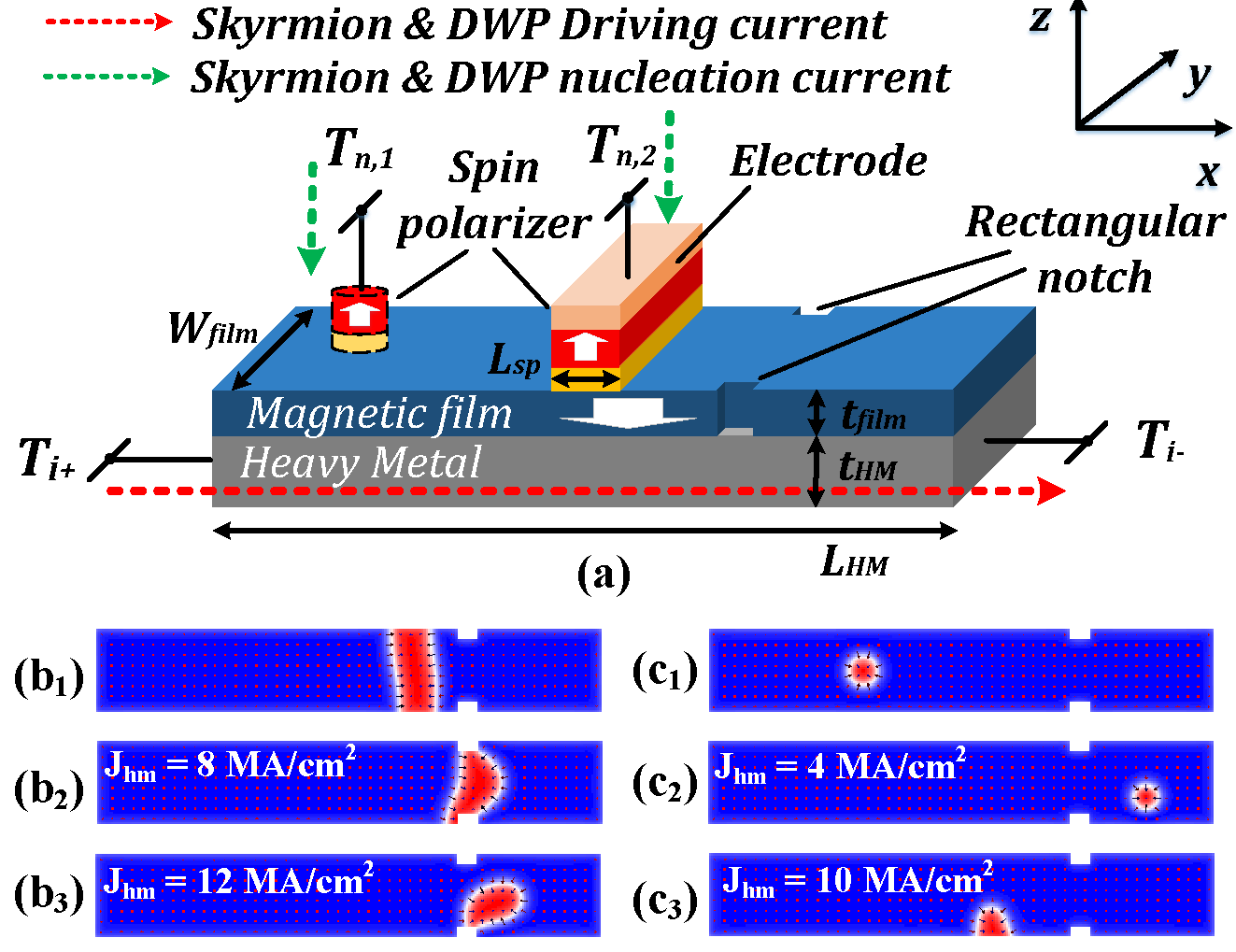}
	\vspace{-.05in}
	\caption{(a) Multilayer device structure with rectangular notch ($30~nm \times 10~nm$) (b) Current-driven Domain Wall Pair (DWP) motion. When driving current density is below the critical depinning current ($J_{hm}\leq 12~MA/cm^2$), DWP is trapped within the notch region. Otherwise, the domain wall pair converts into a fractional skyrmion. (c) Current-driven skyrmion motion in the identical nanotrack, which is free from the notch site. However, skyrmion will collide into boundary with $J_{hm}\geq 9.5 MA/cm^2$.}
	\label{Device_strcuture}
\end{figure}

\begin{figure*}[t]
	\begin{center}
		\begin{tabular}{l}
			\includegraphics [width=0.95\linewidth]{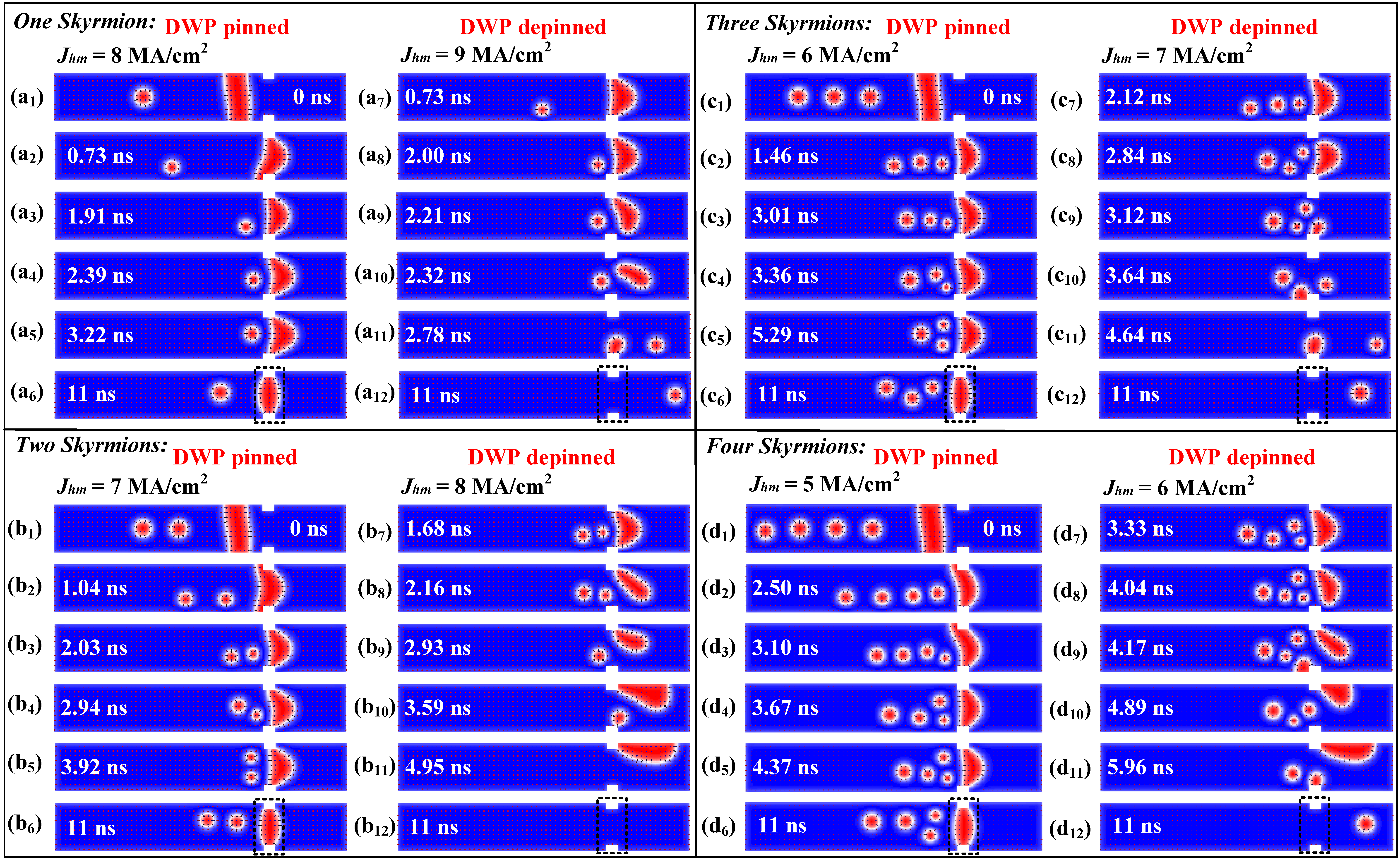}\\
		\end{tabular}
		\caption{Micromagnetic simulation result of dynamic interactions between domain wall pair and skyrmion under 10-ns current pulse, with various number of skyrmions and current densities ($J_{hm}$). For the specific number of skyrmions in the nanotrack, increasing the current density can release the domain wall pair from the pinning site.}
		\label{Dynamics_micromagnetic}
	\end{center}
\end{figure*} 

As the multilayer device structure shown in Fig.~\ref{Device_strcuture}(a), the nanotrack constitutes of heavy metal substrate, ultra-thin ferromagnetic film, and spin polarizers which are mounted on the top for Domain Wall Pair (DWP) and skyrmions nucleation. By vertically injecting current pulse through the spin polarizers \cite{xing2016skyrmion}, DWP and skyrmion can be nucleated within the ferromagnetic film. For the generation of multiple skyrmions, a shift operation is necessary followed by the skyrmion nucleation, which is similar as skyrmion racetrack memory \cite{kang2016voltage}. 
Current-driven skyrmion- and domain wall-motion haven been previously explored as an effective manipulation mechanism to relocate skyrmion and DWP in nanostructures \cite{sampaio2013nucleation}. Compared to directly injecting spin polarized current into magnetic film along x-axis, utilizing Spin Hall Effect (SHE) is a more efficient skyrmion and DWP manipulation method due to higher drifting velocity with identical driving current density \cite{sampaio2013nucleation}. When applying a lateral charge current through heavy metal (T\textsubscript{i+} and T\textsubscript{i-} in Fig.~\ref{Dynamics_micromagnetic}), a pure spin current is generated along z-axis and exerts Slonczewski Spin-Transfer Torque on the spin textures. Such driving force can be described as \cite{sampaio2013nucleation}: 
\begin{equation}
\bm{F}_{STT} = \pm \frac{J_{hm} \theta_{sh} \hbar \pi b}{2e}  \bm{\hat{z}} \times \bm{\hat{\sigma}}
\end{equation}
where $\theta_{sh}$ is the spin Hall angle, $\hbar$ is reduced Planck constant, $b$ is the skyrmion characteristic length, and $ \bm{\hat{\sigma}} $ is the spin polarized direction. The sign of $ \bm{F}_{STT} $ is determined by the sign of topological charge $Q = 1/4\pi \int \bm{m}\cdot (\partial_{x}\bm{m} \times \partial_{y}\bm{m})dxdy$, which is equivalent with the skyrmion number. When the magnetization of one isolated skyrmion core is pointing towards +z, $Q=+1$ whereas $Q=-1$ \cite{jiang2016direct}. It is noteworthy that, all the discussions and mathematical descriptions in the work are based on N{\'e}el-type skyrmion and domain wall, owing to the interfacial DMI \cite{fert2013skyrmions}.

While the driving current is below the critical depinning current density of DWP ($J_{hm} < 12~MA/cm^2$), as seen in Fig.~1(b\textsubscript{2}), the DWP moves along the nanotrack until trapped within the rectangular notch \cite{yuan2014domain}. Otherwise, the DWP converts into a meron \cite{zhou2014reversible}, then annihilates at the nanotrack border when the driving current is turned off. Conversely, as a result of repulsion force between skyrmion and nanotrack boundary, skyrmion has better variation tolerance to move through pinning notch and device edge roughness \cite{sampaio2013nucleation}. However, owing to the skyrmion Hall effect \cite{jiang2016direct} \cite{chen2017spin}, larger driving current density makes skyrmion collides with nanotrack boundary as depicted in Fig.~1(c\textsubscript{3}). Therefore, in this work, we restrict the current density ($J_{hm}$) to be less than $9.5~MA/cm^2$ to assure the correct operation of skyrmions and DWP.

\subsection{Skyrmions dynamics with domain wall pair }
\label{subsec_DWP_Sky}

The skyrmion motion driven by SHE can be modeled by the modified Thiele equation \cite{sampaio2013nucleation} \cite{jiang2016direct} \cite{tomasello2014strategy} \cite{iwasaki2013current} \cite{kang2016skyrmion}:
\begin{equation}
\label{thiele}
\bm{G} \times \bm{v}_{sk} - \alpha \bm{D} \cdot \bm{v}_{sk} + \bm{F}_{STT}+\nabla \bm{V} = 0
\end{equation}
where $\bm{G}$ is the gyro-magnetic coupling vector ($\bm{G} = (0,0,-4\pi Q)$). $\bm{v}_{sk}$ is the velocity of skyrmion motion, which consists of two components $v_{sk,x}$ and $v_{sk,y}$ along x- and y-axis. $\bm{D} = \pi^2 d/(8\sqrt{A_{ex}/K_{u}}) \big [\begin{matrix} 1&0\\ 0&1 \end{matrix} \big ]$ is the dissipative tensor, and $\alpha$ is Gilbert damping factor. $\bm{V}$ is the potential from surrounding environment, such as nanotrack boundary, process impurities and other spin textures \cite{iwasaki2013current}. The four terms in equation.~\ref{thiele} describe Magnus force, dissipative force, Slonczewski in-plane torque and effective Force respectively. Due to the repulsion force of skyrmion-skyrmion, skyrmion-domain wall and skyrmion-edge, the dimension of skyrmion shrinks, which consequently leads to the degradation of skyrmion velocity and driving STT ($ \bm{F}_{STT} $). 

The micromagnetic simulation in Fig.~\ref{Dynamics_micromagnetic} presents the current-driven dynamics between multiple skyrmions and DWP under 10-ns current pulse, with respect to different number of skyrmions and driving current densities ($J_{hm}$). In summary, the simulation results could be classified into two categories based on the pinning or depinning of DWP at the notch.
As depicted in Fig.~\ref{Dynamics_micromagnetic}, for a specific number of skyrmions, domain wall pair can be either pinned or depinned at the rectangular notch by varying driving current density ($J_{hm}$). We take the DWP interacting with four skyrmions as an example. The skyrmions and DWP initially locate at the L.H.S of pinning site at 0ns. Fig.~\ref{Dynamics_micromagnetic}(d\textsubscript{1-6}) shows the skyrmions and DWP dynamics with $J_{hm}=5 MA/cm^2$. Since the DWP is closer to the notch, DWP will first get confined at the pinning site while the skyrmions still move along the nanotrack driven by $ \bm{F}_{STT} $ until approaching the confined DWP. Owing to the mutual repulsion forces between skyrmions and DWP, which is not large enough to overcome the pinning effect, DWP will not be depinned. Then, when the driving current is turned off (after 10-ns), the distorted DWP is relaxed to normal shape. The case of same number (4) of skyrmions with larger current density ($J_{hm}=6~MA/cm^2$) is shown in Fig.~\ref{Dynamics_micromagnetic}(d\textsubscript{1,7-12}), where the DWP gets depinned due to a large enough repulsion force from the accumulated skyrmions. After DWP gets depinned, it converts into a meron and the nanotrack channel is reopened for skyrmions to pass through the notch site. Meanwhile, part of skyrmions will collide with the boundary due to the uneven potentials from notch boundary and other magnetic textures. On the contrary, the skyrmions successfully passing the notch will be moved to the R.H.S, which could be destroyed either by a large current pulse or fork-like device geometry \cite{zhang2014skyrmion}.

\begin{figure}[ht]
	\vspace{0in}
	\centering
	\includegraphics[width=0.40\textwidth]{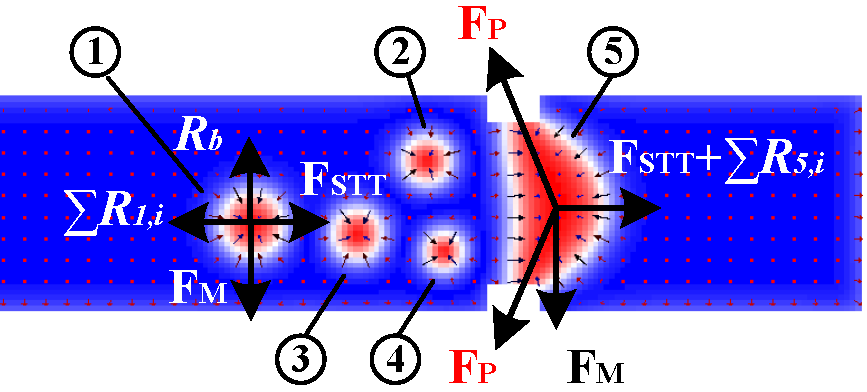}
	\vspace{-.05in}
	\caption{Force analysis of four skyrmions (marked by \textcircled{\raisebox{-0.9pt}{1}} $\sim$ \textcircled{\raisebox{-0.9pt}{4}}) and domain wall pair (marked by \textcircled{\raisebox{-0.9pt}{5}}).   }
	\label{Force_analysis}
	\vspace{-.1in}
\end{figure}

Beyond the general description of skyrmions and DWP interaction, we perfrom the force anaylsis to explain such intriguing dynamics. According to the force analysis shown in Fig.~\ref{Force_analysis}, four skyrmions (\textcircled{\raisebox{-0.9pt}{1}} $\sim$ \textcircled{\raisebox{-0.9pt}{4}}) and DWP (\textcircled{\raisebox{-0.9pt}{5}}) are in the equilibrium state when $J_{hm} = 5~MA/cm^2$. For skyrmion \textcircled{\raisebox{-0.9pt}{1}}, the Magnus force $\bm{F}_m$, skyrmion-boundary repulsion force $\bm{R}_b$, driving STT $\bm{F}_{STT}$ and repulsion force from DWP and other skyrmions $\sum_{i=2}^{5} \bm{R}_{1,i}$ get balanced. Similarly, the forces exerting on DWP in x-axis, which consists of $\bm{F}_{STT}$, x-axis component of confining force $\bm{F}_p$ from pinning site and repulsion force from skyrmions $\sum_{i=1}^{4} \bm{R}_{5,i}$, are in a balanced fashion as well. When $J_{hm}$ is increased to $6~MA/cm^2$, $F_{STT}+\sum_{i=1}^{4} \bm{R}_{5,i}$ increases correspondingly, which is sufficient to overcome the confining force and results in the DWP depinning.

\begin{figure}[ht]
	\centering
	\subfloat[\label{DWP_J_N}]{
		\centering
			\includegraphics[width=0.42\textwidth]{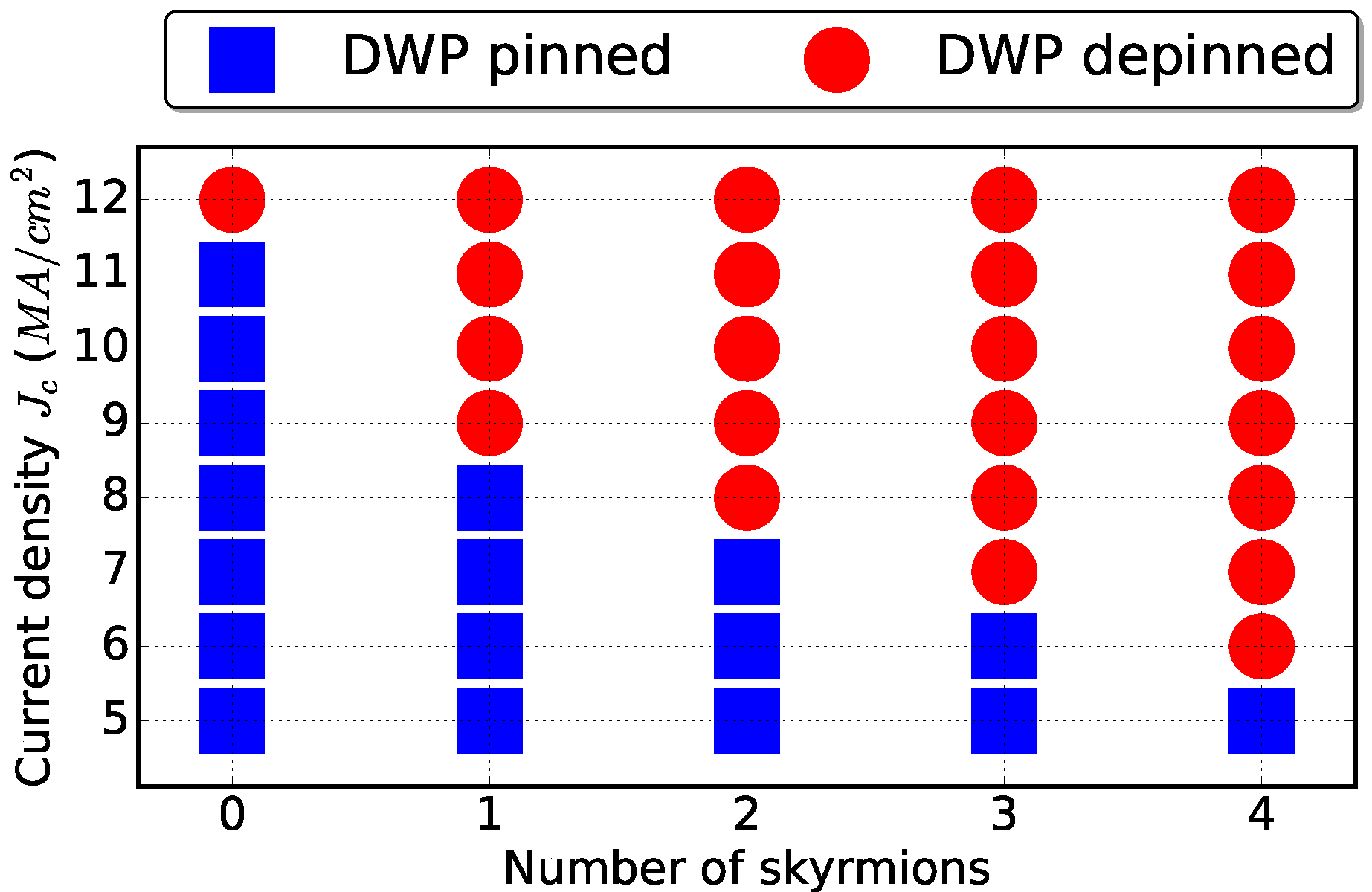}
	} \\
	\vspace{-.05in}
	\subfloat[\label{trade_off}]{
		\centering
	    \includegraphics[width=0.40\textwidth]{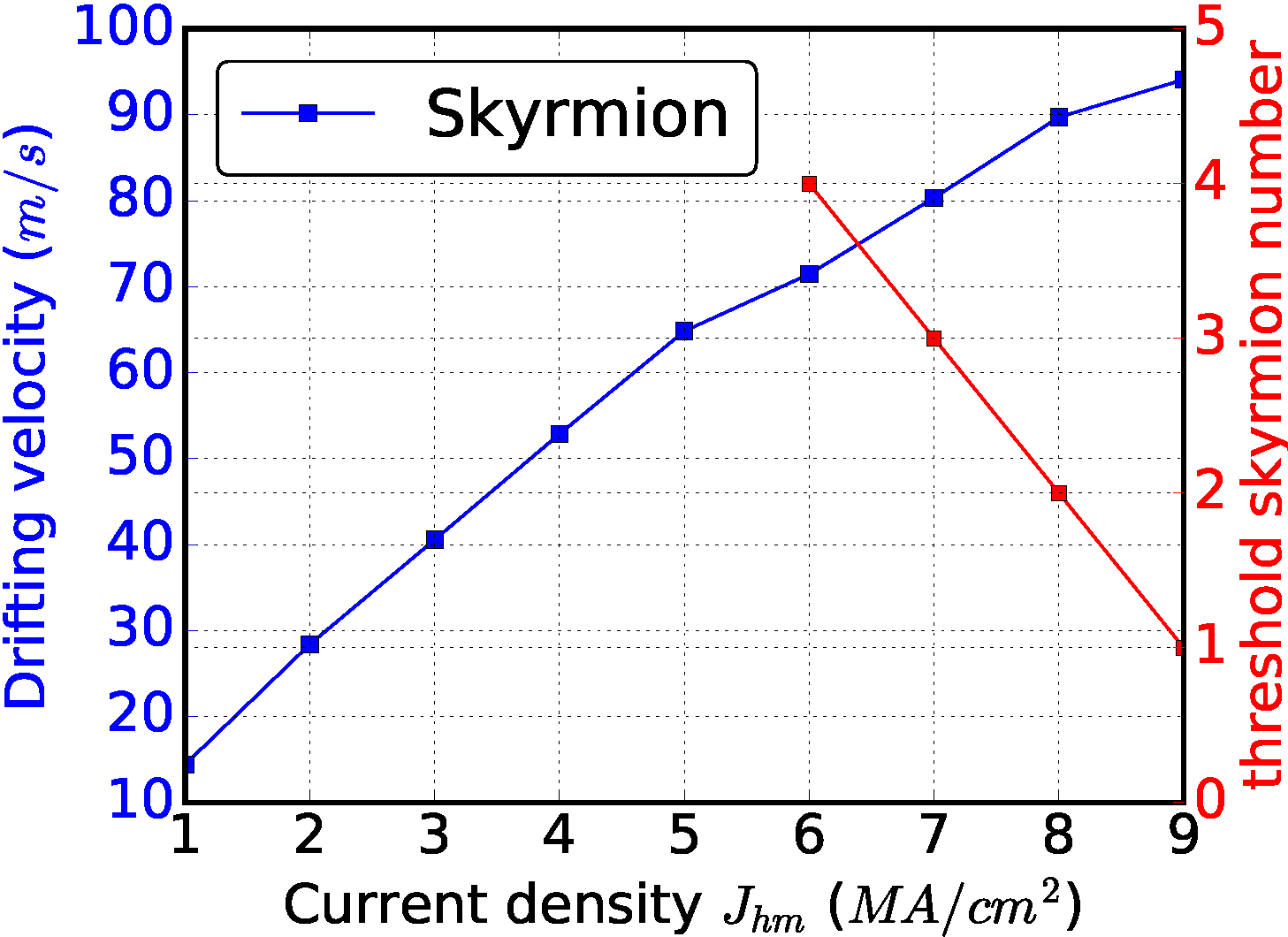}
	} \\
	\vspace{-.05in}
	\caption{(a) DWP pinned/depinned state w.r.t number of skyrmions and current density $J_{hm}$. (b) skyrmion velocity trade off with threshold skyrmion number.}
	\label{data_analysis}
	\vspace{-.1in}
\end{figure}

Such DWP pinning/depinning behavior at the notch determined simultaneously by driving current density and number of skyrmions is summarised in Fig.~\ref{DWP_J_N}. For larger number of skyrmions, the critical DWP depinning current density decreases due to the larger repulsion force from each individual skyrmion. Meanwhile, a lower driving current density requires more skyrmions to depin the DWP out of pinning site, where such quantity of skyrmions is defined as \textit{threshold skyrmion number}. For example, if $J_{hm}=7~MA/cm^2$ and the number of skyrmions is 2, the DWP is pinned. However, considering the same driving current but one more skyrmion in the nanotrack, the DWP is depinned. Therefore, the threshold skyrmion number is 3 for $J_{hm}=7~MA/cm^2$. Moreover, the relationship between skyrmion/DWP drifting velocity, threshold skyrmion number and driving current density is shown in Fig.~\ref{trade_off}. It can be seen that the current-driven skyrmion drifting velocity positively correlates with driving current density, while the threshold skyrmion number reduces with increment of $J_{hm}$. 

\subsection{Device modeling}	
We perform the micromagnetic simulation based on Object Oriented MicroMagnetic Framework (OOMMF). In this work, we focus on the Spin Hall Effect (SHE) induced skyrmion motion. The real-time magnetization dynamics are modeled by Landau-Lifshitz-Gilbert equation with spin transfer torque terms  \cite{donahue1999oommf}:
\begin{multline}
\frac{d\textbf{m}}{dt}=-|\gamma|\textbf{m}\times H_{eff}+\alpha\bigg(\textbf{m}\times \frac{d\textbf{m}}{dt}\bigg) \\ + |\gamma| k (\textbf{m}\times \textbf{m}_p \times \textbf{m}) -|\gamma| k \beta (\textbf{m}\times\textbf{m}_p)
\end{multline}
\begin{equation}
\label{eq_beta}
k = |\dfrac{\hbar}{2\mu_0 e}| \dfrac{J_{hm} \theta_{SH}}{t_{film} M_s}
\end{equation}
where $\hbar$ is the reduced plank constant, $\gamma$ is the gyromagnetic ratio, $t_{film}$ is the thickness of magnetic film, $ \beta $ is the non-adiabatic Spin transfer torque coefficient, and $H_{eff}$ is the effective magnetic field. 

The Dzyaloshinskii-Moriya Interaction (DMI) is applied with DMI extension module  \cite{donahue1999oommf}, which describes the energy density of DMI as \cite{sampaio2013nucleation}:
\begin{equation}
\varepsilon_{DM} = D \bigg(m_z\frac{\partial m_x}{\partial x}-m_x\frac{\partial m_z}{\partial x}+m_z\frac{\partial m_y}{\partial y}-m_y\frac{\partial m_z}{\partial y}\bigg)
\end{equation}
where $D$ is DMI constant, and $m_x$, $m_y$, $m_z$ are the components of normalized magnetization on x-, y- and z-axis respectively.

\begin{table}[ht]
	\caption{Device and simulation parameters} 
	\centering 
	\scalebox{0.92}{
		\begin{tabular}{c c} 
			\hline\hline 
			\rule{0pt}{3ex}
			\textbf{Parameter} & \textbf{Value}  \\
			\hline
			\rule{0pt}{3ex}
			Magnetic film dimension, $(L \times W\times t)_{film}$ & $510\times80\times 0.4$ $nm^3$  \\
			Heavy metal dimension,  $(L \times W\times t)_{hm}$ & $510\times80\times 2$ $nm^3$  \\
			rectangular notch dimension,  $(L \times W)_{notch}$ & $ 30 \times 10 nm^2$ \\
			Gilbert Damping Factor, $\alpha$ & $0.3$  \\
			Non-adiabatic STT factor, $\beta$ & $0.1$ \\
			Spin Hall angle, $\theta_{sh}$ & $ 0.3 $ \\
			Exchange stiffness, $A_{ex}$ & $15\times 10^{-12}$ $J/m$ \\
			Perpendicular Magnetic Anisotropy, $K_u$ & $0.8$ $ MJ/m^3$\\
			Saturation Magnetization, $M_s$ & $5.8\times 10^5$ $A/m$ \\
			DMI constant, $D$ & $3.5$ $mJ/m^2$\\
			Mesh size, $dx\times dy\times dz$ &$2\times2\times0.4$ $nm^3$ \\
			\hline\hline 
		\end{tabular}}
		\label{parameters} 
	\end{table}

\section{Skyrmion-based Majority Logic Gate}

Based on the previous discussion, it can be seen that a threshold-tunable step function could be implemented by such magnetic racetrack due to the dynamical interactions of skyrmions and domain wall pair, in which the skyrmions (presence or absence) are inputs and threshold could be tuned by the driving current density. In this section, a three-input Skyrmion-based Majority Logic Gate (SMLG) is proposed to leverage such unique property. Note that, majority gate is Boolean-complete and is able to implement any Boolean functions \cite{fan2016low}. In the proposed SMLG, the binary input `1' and `0' are represented by the presence and absence of skyrmion, respectively. The output is indicated by the pinned- (`0') or depinned-DWP (`1'). It is noteworthy that it is difficult to sense the existence of skyrmion due to its ultra-small dimension \cite{kang2016skyrmion}, which is one of the main limitations of skyrmion based logic \cite{zhang2015magnetic_logic} and memory designs \cite{kang2017compact}. In our proposed logic design, the output (i.e. domain wall pair) could be easily sensed using a sensing MTJ similar as other domain wall based logic designs \cite{he2016energy}.

\begin{figure}[h]
	\vspace{0in}
	\centering
	\includegraphics[width=0.47\textwidth]{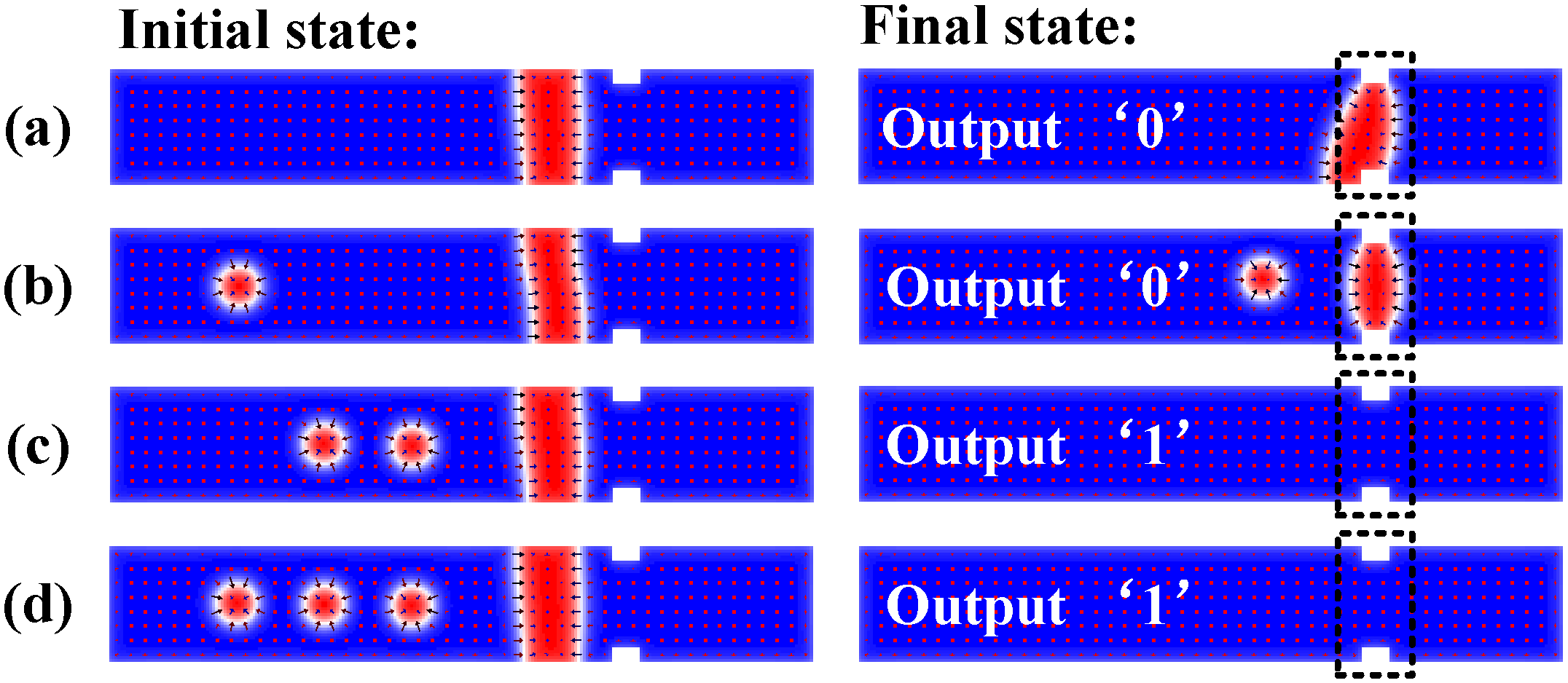}
	\vspace{-.05in}
	\caption{The micromagnetic simulation of skyrmion-based majority gate with different number of skyrmion when $J_{hm}=8~MA/cm^2$. (a-c) shows the input combinations of AND logic, while (b-d) shows the input combinations of OR logic.}
	\label{Logic}
	\vspace{-.1in}
\end{figure}

The output of an n-input Majority Gate (MG) (n is odd) is determined by majority of its inputs. For example, the output is `1' only when more than ($\frac{n-1}{2}$) of the inputs are `1'. 
There are three inputs for the SMLG design, assuming input `A', `B' and `bias' respectively. As reported in the previous section.~\ref{subsec_DWP_Sky}, the threshold skyrmion number is 2 corresponding to the current density $J_{hm}$ of $8~MA/cm^2$. Fig.~\ref{Logic} shows that if the number of input `1's (i.e. number of skyrmions), is greater or equal than 2, the output is `1' (i.e. DWP depinned), otherwise output is `0'. Therefore, such SMLG can also be configured as AND or OR logic through controlling the bias\cite{fan2016low}. The truth table of AND and OR logic is tabulated in Table \ref{SMLG_tab}. 

\begin{table}[ht]
\centering
\caption{Truth table of SMLG for AND/OR logic operation }
\label{SMLG_tab}
\scalebox{0.97}{
\begin{tabular}{|c|c|c|c|c|c|c|c|c|c|}
\hline
\multicolumn{5}{|c|}{AND} & \multicolumn{5}{c|}{OR} \\ \hline
A & B & Bias & DWP & Out & A & B & Bias & DWP & Out \\ \hline
0  &  0 &  0   &  pinned   & 0  &  0 & 0 & 1 &  pinned  &  0   \\ \hline
0  &  1 &  0   &  pinned   & 0  &  0 & 1 & 1 &  depinned &  1   \\ \hline
1  &  0 &  0   &  pinned   & 0  &  1 & 0 & 1 &  depinned  &   1  \\ \hline
1  &  1 &  0   &  depinned & 1  &  1 & 1 & 1 &  depinned   &  1   \\ \hline
\end{tabular}}
\end{table}

Although the presented work employs logic design as a case study, such behavior that accumulation of skyrmions causing the depinning of DWP shows interestingly similar functionality as the Integrate\&Fire neuron of spiking neural network, in which the basic neuron function is to integrate the input synaptic spikes and generate a fire when reaching a threshold. Our future work is to explore such intrinsic match of skyrmion dynamics with neuromorphic computing model, which is promising to implement an ultra-dense and low power brain-inspired spiking neural network.

\section{Conclusion}
In this letter, we present the investigation of dynamic interaction between multiple skyrmions and domain wall pair in a nanotrack. We discover that domain wall pair could be pinned or depinned by the rectangular notch pinning site depending on both the number of skyrmions and driving current density. Such skyrmion and domain wall pair behavior intrinsically performs a threshold-tunable step function taken number of skyrmions as input variable. It can be leveraged to design a Skyrmion-based majority logic gate .

\bibliographystyle{IEEEtran}
\bibliography{IEEEabrv,./bare_jrnl_transmag}

\end{document}